\documentclass[]{aa}  
\usepackage{graphicx}
\usepackage{txfonts}
\def\lappr{\lower 3pt\hbox{$\buildrel < \over \sim\;$}}
\def\llappr{\lower 3pt\hbox{$\buildrel > \over \sim\;$}}

\begin{document}
%

\def\topfraction{.99}
\setcounter{bottomnumber}{1}
\def\bottomfraction{.3}
\setcounter{totalnumber}{3}
\def\textfraction{.01}
\def\floatpagefraction{.9}
\setcounter{dbltopnumber}{2}
\def\dbltopfraction{.7}
\def\dblfloatpagefraction{.5}

  \title{Searching for absorbed AGN in the 2XMM-{\em Newton} pre-release EPIC Serendipitous Source Catalogue$^{*}$}


   \author{E. Memola,
          \inst{1}
	  A. Caccianiga,
	  \inst{1}
	  F. Cocchia,
	  \inst{1}
	  R. Della Ceca,
	  \inst{1}
	  T. Maccacaro,
	  \inst{1}
	  P. Severgnini,
	  \inst{1}
	  D.J. Fyfe,
	  \inst{2}
	  S. Mateos,
	  \inst{2}
	   M.G. Watson,
	   \inst{2}
	   \and
	   G. Lamer
	  \inst{3}
          }

   \offprints{E. Memola \\ email: elisabetta.memola@brera.inaf.it \\ $^*$ Based on observations performed
at the European Southern Observatory, La Silla, Chile and on observations obtained with XMM-{\em Newton},
an ESA science mission with instruments and contributions directly funded by ESA Member States and the USA (NASA).} 

\institute{INAF - Osservatorio Astronomico di Brera, Via Brera 28, 20121 Milano, Italy 
\and  Department of Physics \& Astronomy, University of Leicester, Leicester LE1 7RH, UK
\and AIP - Astrophysikalisches Institut Potsdam, An der Sternwarte 16, 14482 Potsdam, Germany  \\
             }

   \date{Received October 15, 2006; accepted December 29, 2006}

 
\abstract
{}
{We aim to test a method of efficiently selecting X-ray obscured AGN in the 
2XMM-{\em Newton} EPIC Serendipitous Source Catalogue.}
{By means of a strong correlation established using the XMM-{\em Newton} Hard Bright Sample
between the intrinsic absorption and the hardness ratio 
to the 0.5--2.0 keV and 2.0--4.5 keV bands, 
an efficient way of selecting absorbed sources has been worked out. 
A hardness ratio selection based on the 2XMM-{\em Newton} pre-release
Source Catalogue led us to the definition of candidates likely to
be obscured in X-rays.}
{X-ray and optical spectral analysis were performed for three objects.
Strong absorption (N$_{\rm H} > 10^{22}$~cm$^{-2}$) was detected from the X-ray analysis,
confirming the efficiency of the method used to select obscured sources.
The presence of absorption is also revealed in the optical band, although at a 
significantly lower level than inferred from the X-ray band.  
}
{}

\authorrunning 
{E. Memola et al.} 
   \maketitle
%

\section{Introduction}
Medium-deep X-ray surveys carried out with {\em Chandra} and XMM-{\em Newton}
(see \cite{brandt} and references therein for a recent review) 
have shown that obscured active galactic nuclei (AGN) are the main contributors
to the cosmic X-ray background (XRB). \cite{iwasawa} found out
that at least  85\% of the emitted intrinsic
flux of the XRB spectrum is absorbed, by assuming the absorber to be dusty gas.
Therefore, understanding the characteristics of obscured AGN
(i.e.~luminosity and redshift distribution, spectral energy distribution,
the properties of the circumnuclear medium, cosmological evolution)
is a key issue
in unveiling the accretion
process on super massive black holes and tracing and following 
its evolution with cosmic time.

To efficiently select
sizeable and representative samples of absorbed AGN is not an easy task, even at bright fluxes. 
The strong absorption that affects the intrinsic emission of these sources does not allow us
to study, and in several cases even to reveal (e.g.~\cite{rdc4}), their nuclear properties in the optical 
and soft X-ray (E$<$2~keV) bands. Absorbed AGN can be efficiently selected (at least the Compton-thin ones)
and studied using hard X-ray (E$\sim$2--10~keV) bands where the effect of absorption
is less severe and where the host galaxy contribution is minimal. 
Nevertheless, as shown e.g.~by the XMM-{\em Newton} hard bright sample, HBS
(\cite{caccia1}; \cite{rdc1}; Della Ceca et al.~{2007, \em in preparation}) 
at fluxes above $\sim 7 \times 10^{-14}$ erg cm$^{-2}$ s$^{-1}$, 
even in the ``hard'' 4.5--7.5 keV band, absorbed 
AGN\footnote{We use here and in the following N$_{\rm H}=10^{22}$ cm$^{-2}$ as the dividing 
value between unabsorbed and absorbed sources in the X-rays.}
represent only the $\sim25$\% of the sources.
It is therefore essential to increase the efficiency in selecting X-ray absorbed AGN
in order to build up well-defined and sizable samples of these interesting sources. 

Here we introduce a powerful method for selecting bright X-ray absorbed candidates and 
discuss the results of the X-ray and optical spectral analysis for three sources. 
Throughout this paper we adopt
H$_{\rm o}$ = 70 km sec$^{-1}$ Mpc$^{-1}$, $\Omega_{\rm M}$ = 0.3, 
and $\Omega_{\rm \Lambda}$ = 0.7. 
We define a power-law spectrum such that dN/dE = AE$^{-\Gamma}$ where N is the number of photons,
E the photon energy, A the normalization, and ${\Gamma}$ the photon index.

 

\section{Bright, obscured AGN candidates in the 2XMM-{\em Newton} pre-release EPIC Serendipitous Source Catalogue}
\label{par2}

The work presented here is based on the results from the 
XMM-{\em Newton} Hard Bright Sample (\cite{caccia1}; \cite{rdc2}; \cite{rdc1}).
The survey was performed in the 4.5--7.5~keV band
at a flux limit of F$_{\rm x} \sim 7 \times 10^{-14}$ erg cm$^{-2}$ s$^{-1}$. 
The source sample is now almost fully spectroscopically-identified (65$/$67 sources identified),
and the optical and X-ray spectral analysis is complete 
(\cite{rdc1}; Caccianiga et al.~{2007, \em in preparation}; Della Ceca et al.~{2007, \em in preparation}).  
The authors found that X-ray
absorbed and unabsorbed AGN are clearly distinguishable by means of the hardness ratio
HR2, defined as ${\rm HR2}=\frac{CR3-CR2}{CR3+CR2}$, where CR2 and CR3 are the PSF and
vignetting-corrected countrates in the 0.5--2.0 and 2.0--4.5~keV energy band, respectively
(see Fig.~3 in \cite{rdc1}). 
By considering the constraint ${\rm HR2}>0$, at the survey flux limit, 
they found 14 sources out of 15 ($\sim$93$\%$) to have absorbing column densities higher than $\sim$10$^{22}~{\rm cm}^{-2}$,
and 4 of these sources ($\sim$30\%) turned out to be the rare X-ray type~2 quasars 
(L$_{\rm x} > 10^{44}$ erg s$^{-1}$; N$_{\rm H} > 10^{22}~{\rm cm}^{-2}$).

The XMM-{\em Newton} Survey Science Centre (SSC)\footnote{The XMM-Newton Survey Science Centre (SSC,  see
http://xmmssc-www.star.le.ac.uk) is an international collaboration, involving a
consortium of 10 institutions, appointed by ESA to help the SOC in developing
the software analysis system, to pipeline  process all the XMM-{\em Newton} data, and 
to exploit XMM-{\em Newton} serendipitous detections.} 
is assembling 
the second XMM-{\em Newton} EPIC Serendipitous Source Catalogue
(Watson et al.~{2007, \em{in preparation}}).
A first preliminary version of the 2XMM Serendipitous EPIC Source Catalogue (2XMMp) 
has been recently released and is available at http://xmm.vilspa.esa.es/xsa/.
The catalogue contains 153105 X-ray source detections 
drawn from 2400 useful XMM-{\em Newton} EPIC public observations. 
The median flux (in the total photon energy band 0.2$-$12 keV) of the 
catalogue sources is $\sim2.4 \times 10^{-14}$ erg cm$^{-2}$ s$^{-1}$; 
$\sim$20\% of the sources having fluxes below $10^{-14}$ erg cm$^{-2}$ s$^{-1}$. 

The values of HR given in the catalogue are defined using energy bands 
that are slightly different from the ones used in the HBS. The HR closer to the HR2 used
in the HBS (see above) is the HR3, defined as ${\rm HR3}=\frac{CR4-CR3}{CR4+CR3}$, 
where CR3 and CR4 are the PSF and vignetting-corrected countrates in the 
1.0$-$2.0 and 2.0$-$4.5~keV energy band, respectively.
In the 2XMMp catalogue, there are 427 high latitude ($|b|>20^{\circ}$)
sources, which have HR3\footnote{HR3 refers here to
MOS2 data, which were also used for the HBS.} $>$ 0
and MOS2 detection likelihood (ML)
in the 4.5--12 keV energy band $>$ 20 (\cite{cash}).
This group of sources represents a crucial starting point 
for testing the efficiency in selecting absorbed AGN.
We picked up three bright X-ray sources with
F$_{\rm x} > 10^{-13}$ erg cm$^{-2}$ s$^{-1}$ in the energy band 4.5--12~keV
and a bright optical counterpart (R $<$ 17~mag)
to analyse their X-ray and optical spectra.
The optical counterparts were observed 
at the 4-m New Technology Telescope (NTT@ESO) in Chile last March 2006,
as part of a more general programme concerning the Hard Bright Sample.
In the following we present and discuss our results.
\section{X-ray spectral analysis}
\label{par3}

We performed the X-ray spectral analysis of XMM-{\em Newton}
data for the following sources:
2XMMpJ115121.8$-$283605, 2XMMpJ145442.2+182937, and 2XMMpJ151612.2+070341.
They were observed between August 2000 
and February 2003 with exposure times ranging from $\sim$~23 up to $\sim$~42 ksec
(for the observation details see Table \ref{xmmlog}).
The EPIC MOS1, MOS2, and pn detectors were operated in Prime Full Window Mode.  

All MOS and pn data were reprocessed with the XMM-{\em Newton}
Science Analysis Software (XMM-SAS) version 6.5.0,
using the latest calibration products.
Source counts for MOS1, MOS2, and pn (when available) 
were extracted in the 0.2--12~keV band
from a circular region, centred on the source, 
which had a radius of 25 arcsec for 2XMMpJ115121.8$-$283605,
14 arcsec for 2XMMpJ145442.2+182937, and 12 arcsec for 2XMMpJ151612.2+070341,
depending on the countrate statistics and source off-axis.
Background spectra were extracted from larger
source-free circular regions (radius of $\sim$ 40$-$80 arcsec) close to the object.
We selected single and double events (PATTERN$\leq$4) for the pn
and single, double, triple, and quadruple events (PATTERN$\leq$12) 
for the MOS.
Our analysis is based on both EPIC MOS (MOS1 and MOS2) and pn data.
In particular, for both MOS1 and MOS2,
we extracted source and background spectra
and then summed them up {\em a posteriori} by means of {\em mathpha},
ending up with one source spectrum and one background spectrum
region valid for a {\em global} MOS1+MOS2 case.
The ancillary response matrix (ARF) and the detector response matrix (RMF)
were created by the XMM-SAS tasks {\em arfgen} and {\em rmfgen}.
The {\em arfgen} routine convolves the mirror effective area (vignetting modified), filter 
transmission, and CCD efficiencies as a function of energy to return the ARF component of response.
The ARF, at the source position on the detector,
is emission-weighted, and the weighting falls off with distance from the source in line with the PSF.
The RMF is spatially averaged using an appropriate PSF detector map.
%
%
%
%
The response matrixes were also merged by using {\em addrmf} and {\em addarf}.
Due to the countrate of our sources (see Table \ref{xmmlog}), pile up is not a problem.

We used the XSPEC package (version 11.3.1; \cite{arnaud})
in order to perform the spectral fitting analysis.
MOS and pn spectra were fitted simultaneously
(when both data were available) and respectively
in the 0.3--10 (2XMMpJ115121.8$-$283605), 0.7--10 (2XMMpJ145442.2+182937), 
and 0.5--10 (2XMMpJ151612.2+070341) keV energy range.
The source counts were grouped
into energy bins such that each bin contains 
at least 15 counts where the fit statistic in use was chi-squared ($\chi^2$).
The quoted errors on the best-fit parameters correspond to the 90\% confidence level
for one interesting parameter (i.~e., $\Delta\chi^2 = 2.71$; \cite{avni}).
To analyse the X-ray data we made use of the redshifts 
derived from the optical spectra
(see $\S$\ref{par4} and Table \ref{xmmtable}). 


\subsection{2XMMpJ115121.8$-$283605} 

\begin{figure}
   \centering
\includegraphics[width=6.3cm, angle=-90]{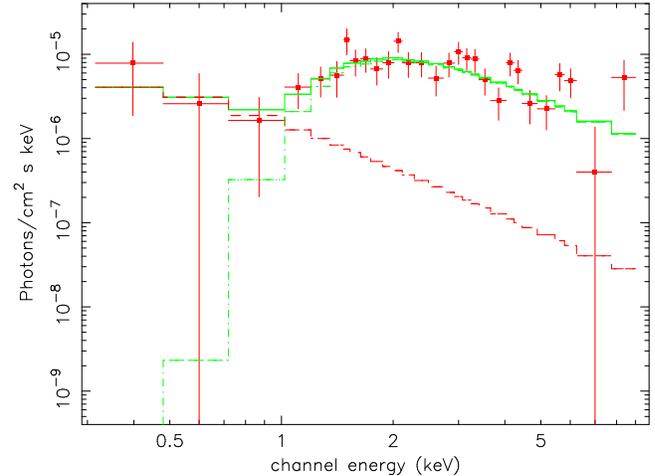}
     \caption{2XMMpJ115121.8-283505: XMM-{\em Newton} MOS folded spectrum.
The best-fit model is a leaky-absorbed power-law.
              }
         \label{X-1151}
   \end{figure}

The source 2XMMpJ115121.8$-$283605 (see Fig.~\ref{X-1151}) was observed by XMM-{\em Newton}
on January 3, 2002 for about 40~ksec of good exposure time. 
The pn data are not available
because the source is outside the pn field of view.
Fitting the data with a single absorbed power-law model 
leads to obvious residuals in the low-energy range.
Our best-fit model ($\chi^{2}/{\rm dof}=27.41/25$) is therefore based on
a leaky-absorbed power-law continuum (see Table~\ref{xmmtable}), 
where two power-laws are invoked to explain both 
the soft (0.3$-$1.0~keV - {\em unabsorbed} power-law) and the
hard (up to 10~keV - {\em absorbed} power-law) part of the X-ray spectrum.
The leaky-absorbed power-law continuum has a functional form, which provides a simple
and convenient parametrization of the relative contributions of both the primary absorbed
component and the component that produces the flux excess at soft energies. 
The model might represent the physical case of an X-ray source observed both
directly through the absorbing torus (absorbed power-law) and after the 
scattering on a warm, highly ionized gas located outside the
absorbing medium (unabsorbed power-law). Another possible interpretation is the presence of a 
cloudy-absorbing medium, which would be responsible for both the absorbed and the
transmitted component.  
Nevertheless, emission lines at soft X-rays might be present, due to nuclear processes or related
to starburst activity, as well as a thermal component because of the presence of a hot plasma.
We can clearly not prove this hypothesis due to the lack of soft energy statistics.     
The two power-laws are constrained to have the same photon index ($\Gamma \sim 1.9$), whereas 
the normalization of the unabsorbed component results in a few percent ($\sim$3\%) 
of the normalization of the absorbed one, in agreement with previous results on absorbed AGN
(e.g.~\cite{turner}; \cite{rdc3}; \cite{caccia1}). 
The absorbed power-law reveals an intrinsic hydrogen column density, 
N$_{\rm H}$, of about 2.6$\times 10^{22}{\rm cm}^{-2}$, which 
clearly points to an X-ray absorbed source (even with a single-absorbed power-law model
the N$_{\rm H}$ was about 2$\times 10^{22}{\rm cm}^{-2}$). 
The intrinsic X-ray luminosity 
of this source (L$_{\rm x}\sim 1.8\times10^{43}$~erg sec$^{-1}$)
is typical of a Seyfert galaxy.      

\subsection{2XMMpJ145442.2+182937}

\begin{figure}
   \centering
\includegraphics[width=6.3cm, angle=-90]{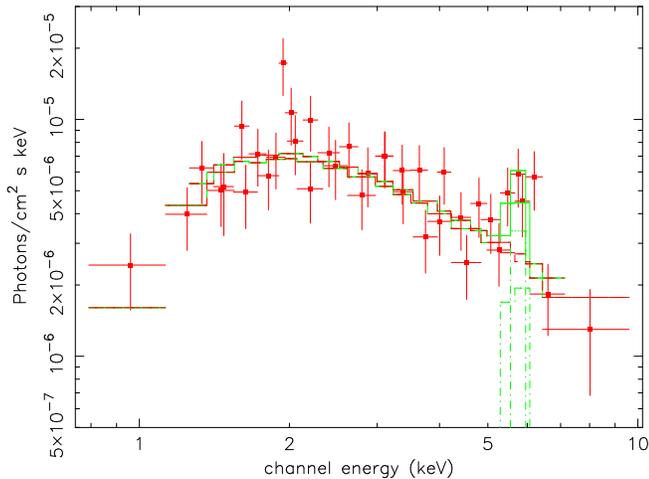}
      \caption{2XMMpJ145442.2+182937: XMM-{\em Newton} MOS+pn folded spectrum.
The best-fit model is a single-absorbed power-law with a Gaussian line.
              }
         \label{X-1454}
   \end{figure}

The source 2XMMpJ145442.2+182937 (see Fig.~\ref{X-1454}) was observed by XMM-{\em Newton} on February 16, 2003
for about 37~ksec of good exposure time by the MOS cameras and 25 ksec by the pn camera. 
The best-fit model ($\chi^{2}/{\rm dof}=30.73/33$) 
based on both MOS and pn data is a single absorbed
power-law along with a Gaussian line detection at 90\% confidence level
(see Table \ref{xmmtable}).
The power-law shows a photon index $\Gamma \sim 1.4$, together with 
N$_{\rm H} \sim 1.5\times 10^{22}~{\rm cm}^{-2}$, which 
points to an X-ray absorbed source. If we freeze the photon index to $\Gamma \sim 1.9$,
the N$_{\rm H}$ value remains quite stable around $\sim 2.4\pm0.4\times10^{22}$~${\rm cm}^{-2}$.
In both cases the iron line equivalent width is about 700~eV with associate errors of about 60$\%$.
This equivalent width value, due to the large error associated to it, is consistent with 
the  value of N$_{\rm H}$ measured from the cut-off. Moreover,  
a detailed analysis of the line properties is not warranted by the quality of the data,
therefore we also can not exclude a possible blending of two narrow iron lines at 6.4 and 6.7~keV.
The X-ray luminosity (L$_{\rm x}\sim 0.7\times10^{43}$~erg s$^{-1}$) is typical of a Seyfert galaxy. \\
%
%

\subsection{2XMMpJ151612.2+070341}

\begin{figure}
   \centering
\hspace*{-0.8cm}
\includegraphics[width=6.6cm, angle=-90]{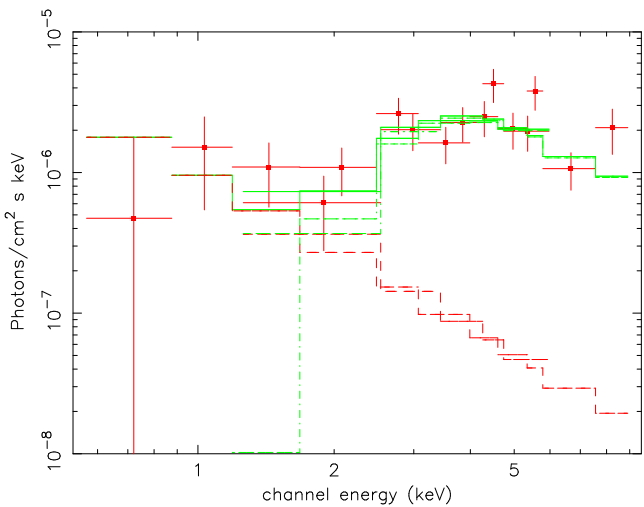}
     \caption{2XMMpJ151612.2+070341: XMM-{\em Newton} MOS+pn folded spectrum.
The best-fit model is a leaky-absorbed power-law.
              }
         \label{X-1516}
   \end{figure}

The source 2XMMpJ151612.2+070341 (see Fig.~\ref{X-1516}) was observed by XMM-{\em Newton} on August 21, 2000
for about 30~ksec of good exposure time with the MOS cameras and 23 ksec with the pn camera.
A single-absorbed power-law  model ($\chi^{2}/{\rm dof}=18.98/12$)
provided a best-fit value of the photon index that was very low
($\Gamma \sim 0.5$),
whereas the value of N$_{\rm H}$ was already high and about $\sim 4\times10^{22}{\rm cm}^{-2}$.
Such a flat $\Gamma$ value would suggest Compton thickness, a thesis otherwise
not supported by the optical data.
After freezing the value of $\Gamma$ to 1.9, more typical of AGN, there were residuals at 1--2~keV. 
The best-fit model ($\chi^{2}/{\rm dof}=16.55/12$) is a leaky-absorbed power-law continuum
with a frozen photon index of 1.9 (see Table \ref{xmmtable}). The statistics do not allow us to leave this
parameter free. The hydrogen column density is N$_{\rm H}\sim 1.4\times10^{23}{\rm cm}^{-2}$,
which points to an X-ray absorbed object.
The X-ray luminosity (L$_{\rm x}\sim 1.6\times10^{43}$~erg s$^{-1}$) is typical of a Seyfert galaxy.

%

\begin{table*}
\begin{center}
\caption{XMM-{\em Newton} observation log.}
\label{xmmlog}
\begin{tabular}{lcccccccc}
\hline \hline
Source  &  Obsid & Obs.Date
        & mos1-exp (s) &  mos2-exp (s) & mos1+mos2-cts       & pn-exp (s)    & pn-cts & Filter$^{*}$ \\
\hline
2XMMpJ115121.8$-$283605      & 0027340101  &   2002 01 03        &  42062   & 37253  & 299  &  no pn  & no pn  & Thin1 \\
2XMMpJ145442.2+182937      & 0145020101  &   2003 02 16        &  36975   & 36421  & 280  &  25535  & 274  & Thin1 \\
2XMMpJ151612.2+070341      & 0109920101  &   2000 08 21        &  30076   & 30123  &  84  &  23085  & 133  & Thin1 \\
\hline 
\end{tabular}
\end{center}
$^{(*)}$ Filter type refers to all the EPIC instruments involved in the corresponding observation.
\end{table*}

\begin{table*}

\begin{center}
\caption{XMM-{\em Newton} spectral analysis: best-fit parameters.}
\label{xmmtable}
\begin{tabular}{lcccccccccc}
\hline \hline
Source & R & z& N$^{(a)}_{{\rm H}_{\rm Gal}}$ & N$_{\rm H}^{\rm(b)}$ & $\Gamma$ & E$_{{\rm Fe}-K\alpha}$ & EW$_{{\rm Fe}-K\alpha}$ & $\chi^{2}/{\rm dof}$  & Flux-mos$^{\rm(c)}$             
& L$^{\rm(d)}$ \\
     & {\scriptsize mag} & & &  &        & keV                     & {\rm eV}&   && \\
\hline
2XMMpJ115121.8$-$283605$^{(1)}$ & 16.64 & 0.177&0.06 &2.6$^{+1.4}_{-1.2}$ & 1.9$\pm{0.5}$ &  --  & -- & $27.41/25$ & 1.76 & 1.8 \\
2XMMpJ145442.2+182937$^{(2)}$    & 16.12 & 0.116& 0.02 & 1.5$^{+0.8}_{-0.5}$ & 1.4$\pm{0.3}$ & 6.34$^{+0.44}_{-0.11}$ & 700$^{+300}_{-480}$ &  $30.73/33 $  & 2.08  & 0.7 \\
2XMMpJ151612.2+070341$^{(1)}$    & 17.20 & 0.176& 0.03 & 14$^{+7.4}_{-4.4}$ & 1.9$^*$ & -- & -- &  $16.55/12 $  & 1.07 & 1.6  \\
\hline
\end{tabular}
\end{center}
$^{(1)}$ Model: Leaky-absorbed  power-law;
$^{(2)}$ Model: Single-absorbed power-law + Gaussian line;
$^{(*)}$ Frozen;
$^{(a)}$ Galactic column density in units of $10^{22}$cm$^{-2}$;
$^{(b)}$ Absorbing column density in the source direction in units of $10^{22}$cm$^{-2}$;
$^{(c)}$ X-ray fluxes in the band 2-10~keV in units of $10^{-13}$ ergs cm$^{-2}$ sec$^{-1}$
and corrected only for the Galactic absorption;
$^{(d)}$ X-ray luminosities in the band 2-10~keV in units of $10^{43}$ ergs sec$^{-1}$ 
and corrected for Galactic and intrinsic absorption.
\end{table*}

%

\begin{figure*}
\vspace{-0.5cm}
\hspace{1.0cm}
\includegraphics[width=6.0cm]{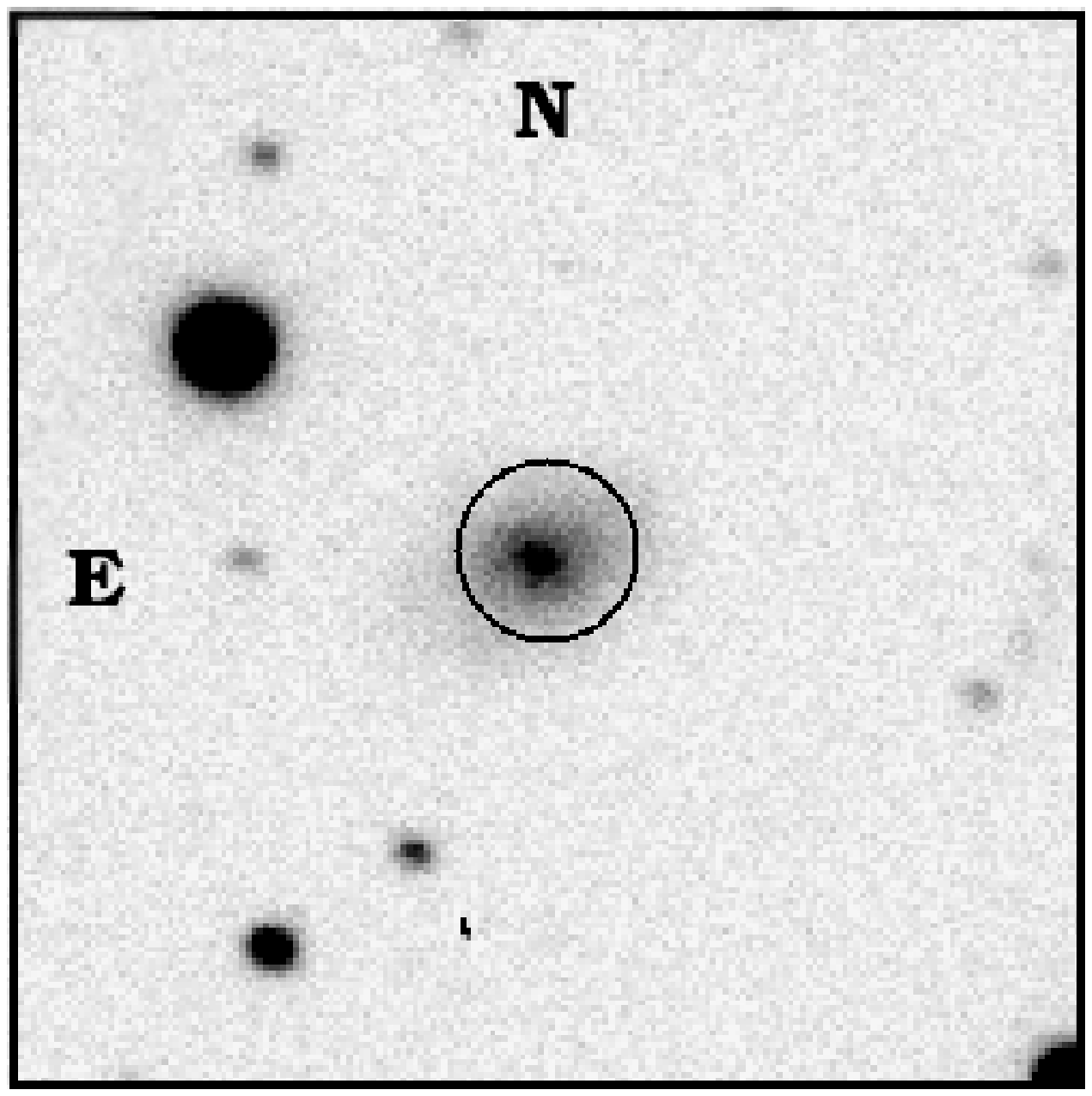}
\hspace{3.0cm}
\includegraphics[width=6.5cm]{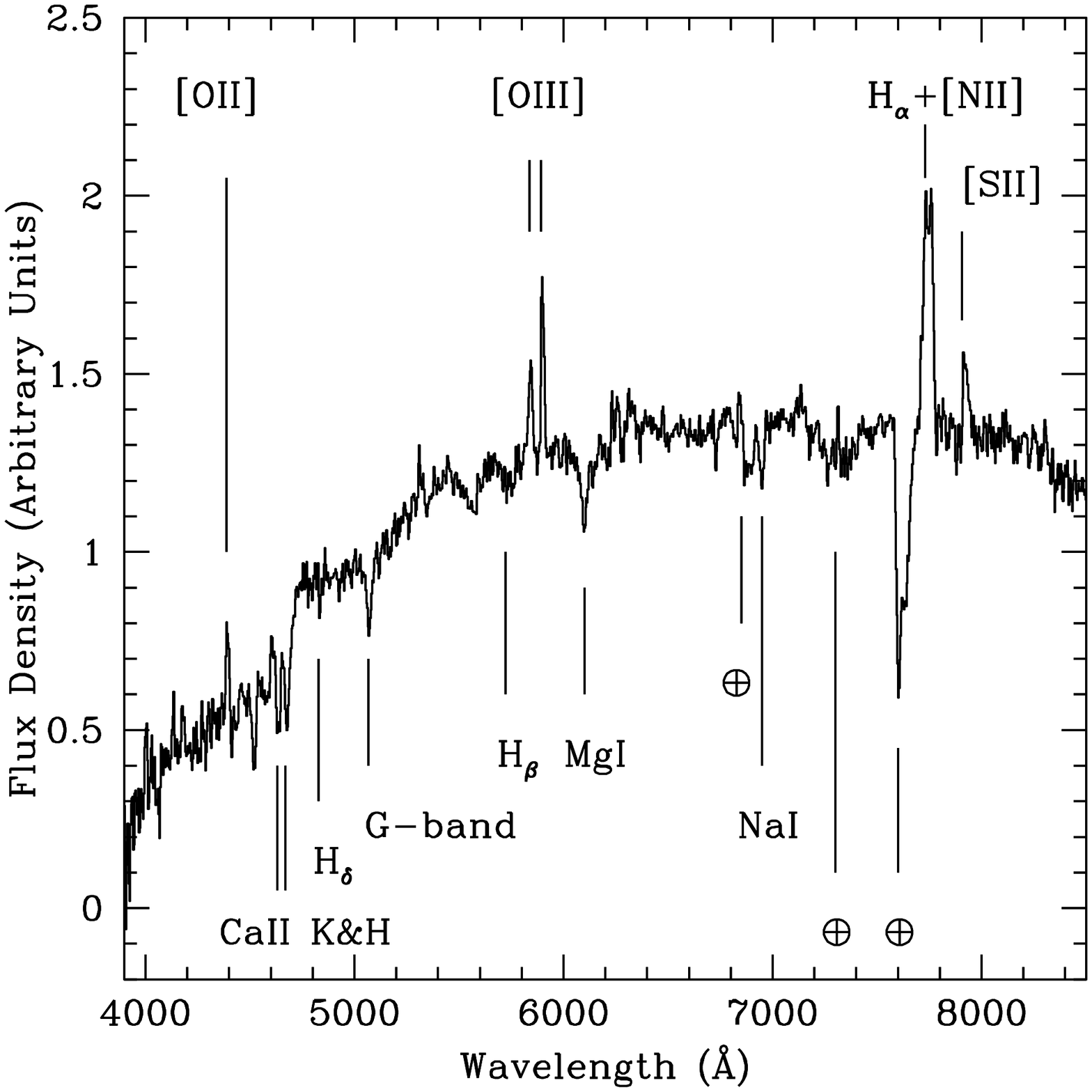}
\caption{2XMMpJ115121.8$-$283605: NTT@ESO R-image (1'$\times$1'). North is on top and east to the left.
A circle of 5'' radius, corresponding to the 95\% confidence level X-ray error circle (see \cite{rdc2}), 
is shown ({\em left panel}) to clearly mark the optical counterpart of the XMM-{\em Newton} source; 
NTT@ESO optical spectrum {\em (right panel)}.}
\label{0-1151}
\end{figure*}

\begin{figure*}
\vspace{-0.5cm}
\hspace{1.0cm}
\includegraphics[width=6.0cm]{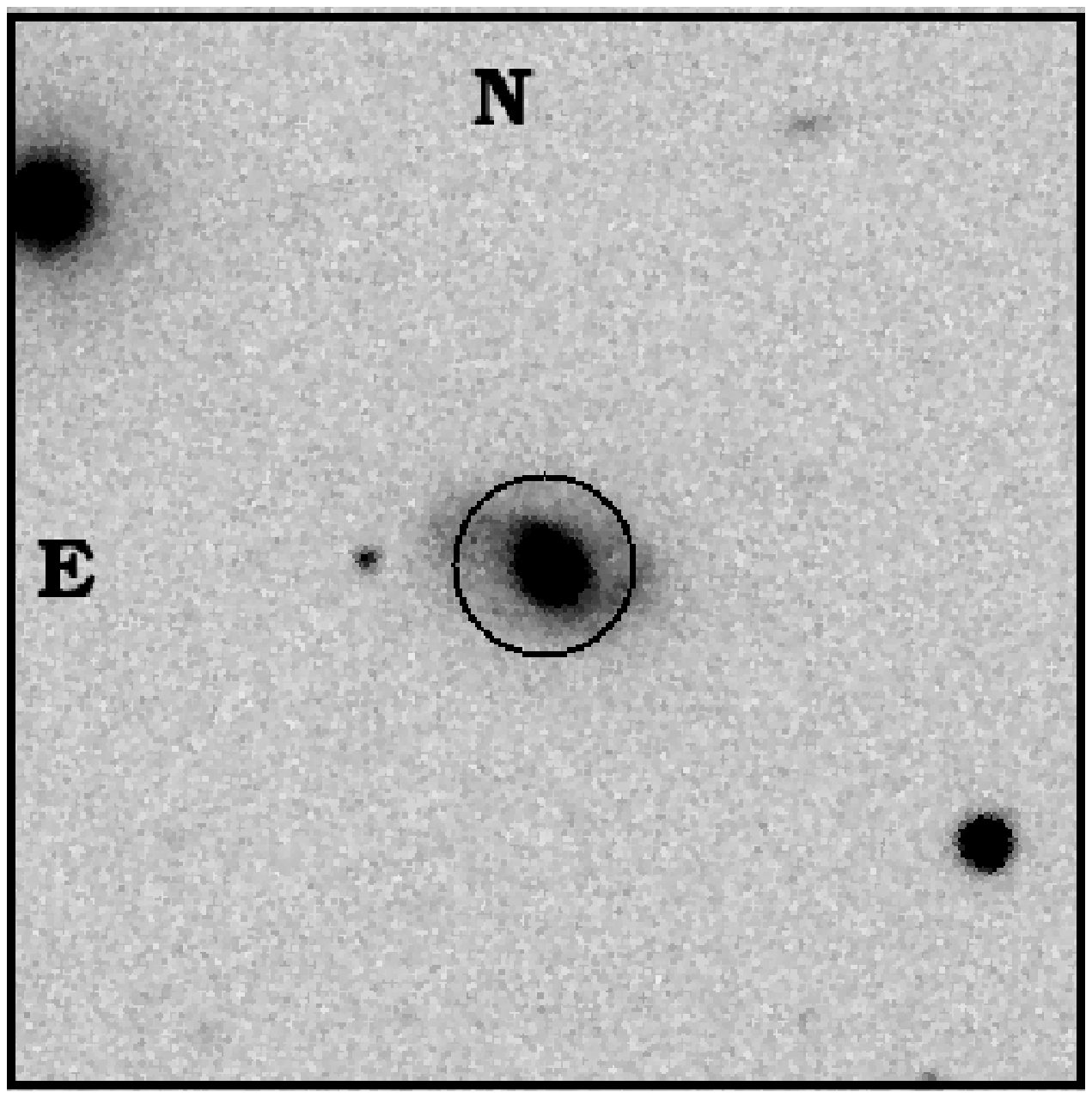}
\hspace{3.0cm}
\includegraphics[width=6.5cm]{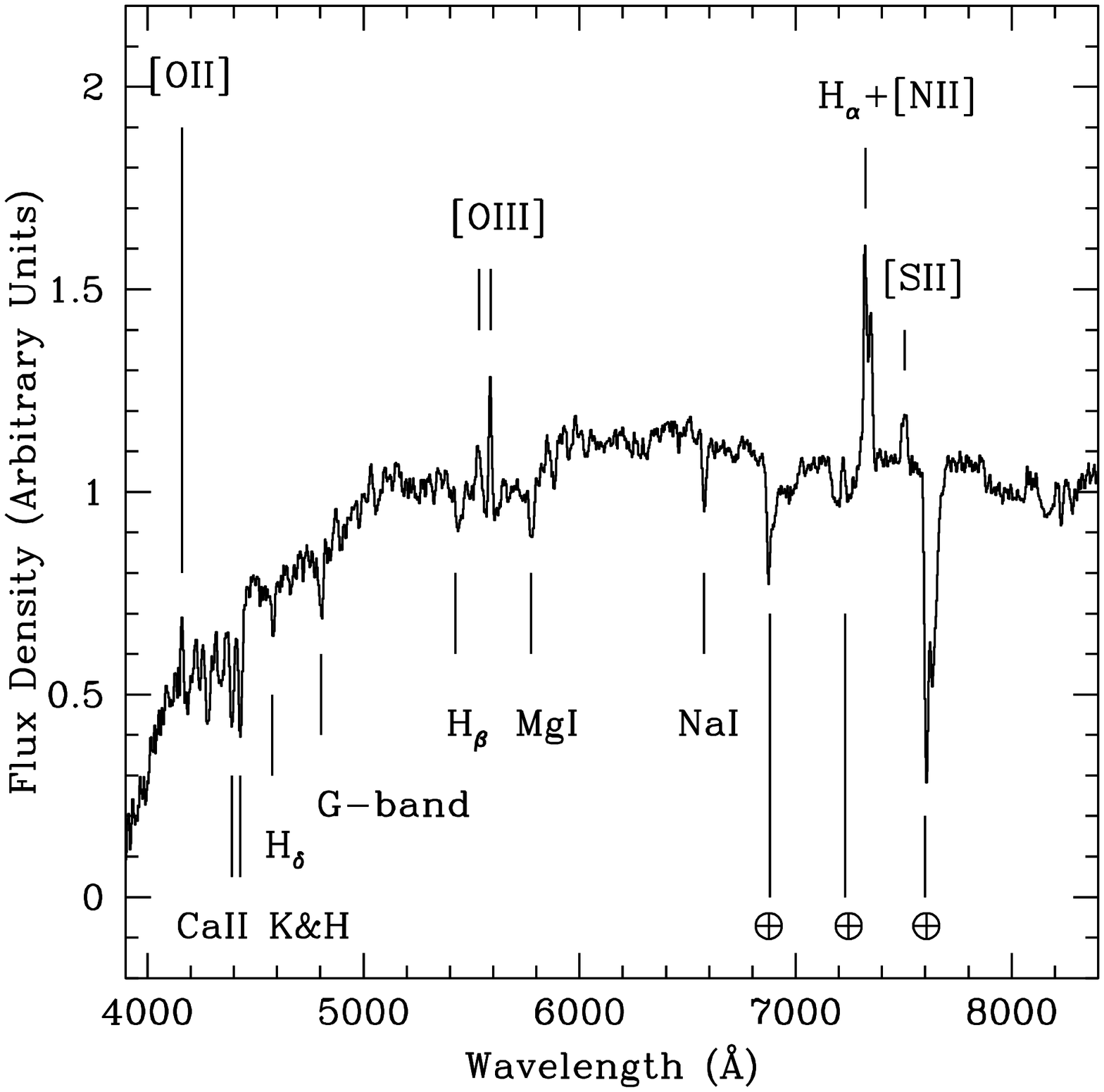}
\caption{2XMMpJ145442.2+182937: NTT@ESO R-image (1'$\times$1').
North is on top and east to the left. A circle of 5'' radius,
corresponding to the 95\% confidence level X-ray error circle (see \cite{rdc2}),
is shown 
({\em left panel}) to clearly mark the optical counterpart of the XMM-{\em Newton} source; 
NTT@ESO optical spectrum {\em (right panel)}.}
\label{0-1454}
\end{figure*}

\begin{figure*}
\vspace{-0.5cm}
\hspace{1.0cm}
\includegraphics[width=6.0cm]{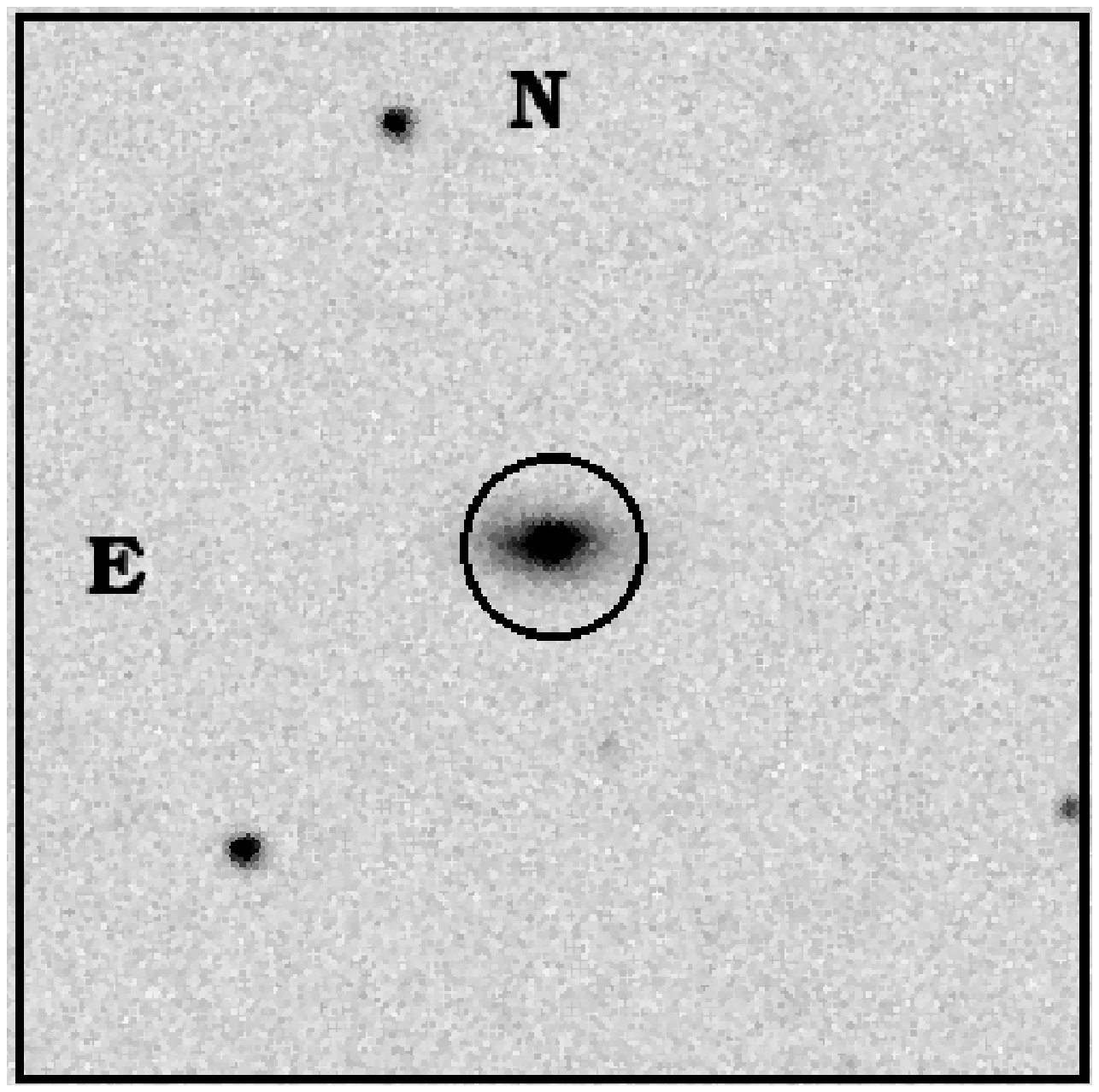}
\hspace{3.0cm}
\includegraphics[width=6.5cm]{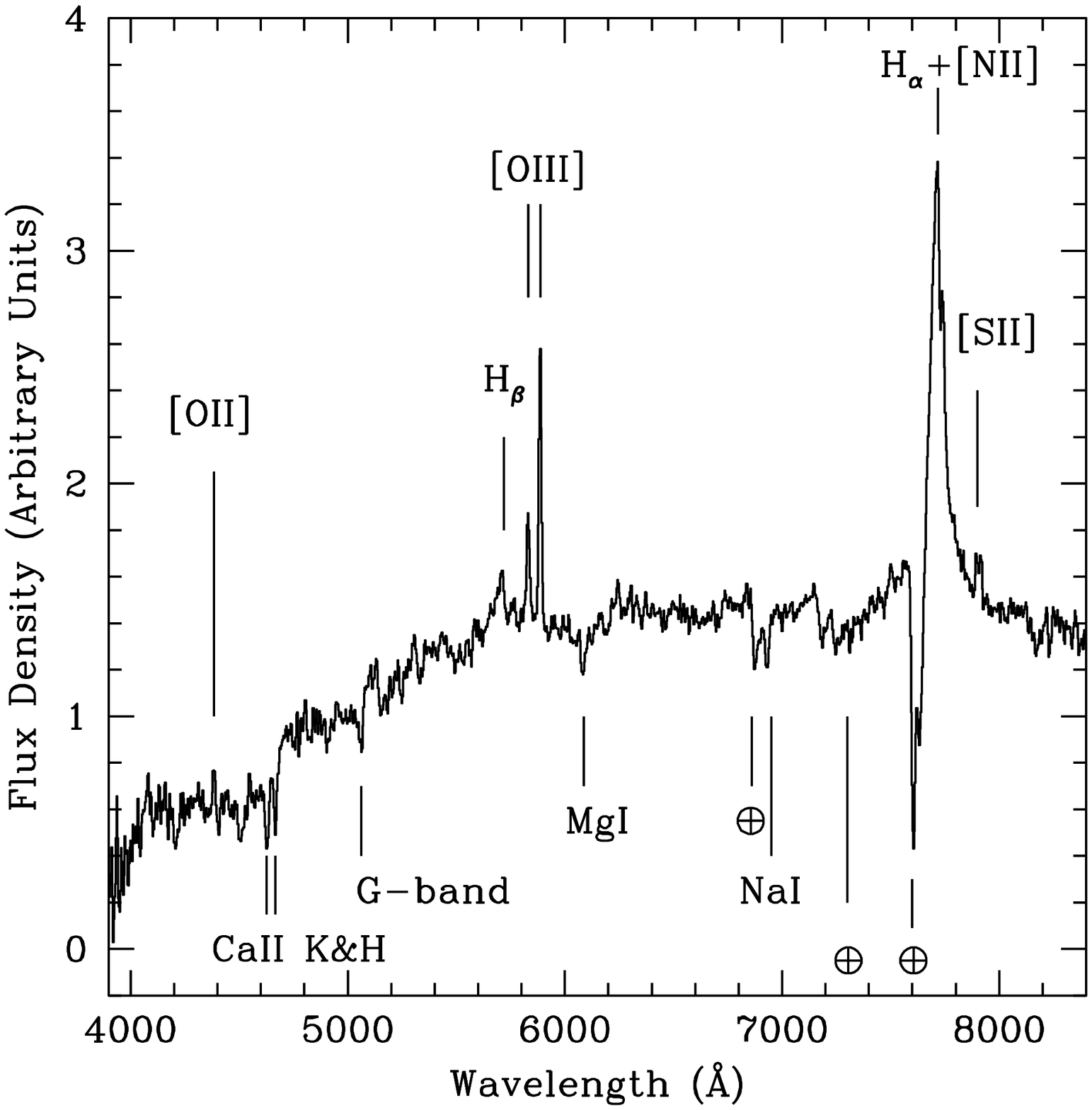}
\caption{2XMMpJ151612.2+070341: NTT@ESO R-image (1'$\times$1').
North is on top and east to the left.
A circle of 5'' radius, corresponding to the 95\% confidence level X-ray error circle (see \cite{rdc2}),
is shown 
({\em left panel}) to clearly mark the optical counterpart of the XMM-{\em Newton} source; 
NTT@ESO optical spectrum {\em (right panel)}.}
\label{0-1516}
\end{figure*}

\section{Optical spectral analysis}
\label{par4}
We performed the spectral analysis of the optical counterparts for the three
X--ray sources.
Optical images and spectroscopy were acquired at the NTT@ESO telescope 
using EMMI. 
We obtained shallow optical images with a typical exposure time of 60--90s. 
Optical images were bias-subtracted, flat-field divided, and 
flux-calibrated using observations of standard stars obtained during each
night. Astrometric calibration was performed using the software package GAIA 
(version 2.6-9 by P.W.~Draper) by matching sources found in the USNO catalogue 
(Monet et al.~1998). We show the R-images of our sources in 
Figs.~\ref{0-1151}, \ref{0-1454}, and \ref{0-1516} ({\em left panels})
and report their R-magnitudes in Table~\ref{xmmtable}.

Long-slit spectroscopy was carried out in the 3800$-$9000 \AA\  
band with a low dispersion ($\sim1.7$~$\AA/{\rm pixel}$), long-slit configuration and 
with a low-resolution grism (Grism$\#$2, R$\sim$600). The seeing during the observing 
run ranged from 1'' to 2'', and we used a slit width of 1''--1.5''. 
Data reduction was performed using  
IRAF\footnote{IRAF is distributed by the National Optical Astronomy  
Observatory, which is operated by the Association of Universities  
for Research in Astronomy, Inc., under cooperative agreement with the  
National Science Foundation.} standard packages.  Wavelength calibrations were  
carried out by comparison with exposures of He-Ar lamps. 
The relative flux calibration of the spectra was obtained using  
observations of spectro-photometric standard stars   
performed within a few hours of the spectroscopy of our sources; 
no attempt was made to obtain an absolute flux calibration of the spectra.

The optical spectra of the three objects are presented in 
Figs.~\ref{0-1151}, \ref{0-1454}, and \ref{0-1516}, ({\em right panels}).
The spectra show relatively strong emission lines, like the 
[OIII]$\lambda$5007\AA\ and
the H$\alpha$-[NII] blend, and a continuum that is highly contaminated (or 
dominated)
by the light of the host galaxy. 
The presence of broad (FWHM$\sim$1900$-$3000 km/s) emission lines is clear in
one source (2XMMpJ151612.2+070341) and is also suggested in 2XMMpJ115121.8$-$283605, 
where the
H$\alpha$ is likely to have a broad component. 
In 2XMMpJ145442.2+182937, instead, only narrow lines are visible in the spectrum.

To estimate the level of optical absorption in the three sources, we 
made use of the method described in Caccianiga et al.~{(2007, \em in preparation)}.
They apply this procedure to studying and classifying the
objects in the XMM-{\it Newton} Bright Serendipitous Survey that are 
contaminated by the {\em host galaxy} light (``diluted'' sources).
In synthesis, this approach involves a spectral model composed of a galaxy plus
an AGN template (\cite{paola}). 
The AGN template, in particular,  has a component  (the
continuum and the broad emission lines) that can be absorbed according to an 
appropriate extinction curve and a second component (the narrow emission 
lines) that is not affected by the absorption. We estimate the optical 
absorption as the  value that best emulates the observed spectrum (both 
the continuum and the emission-line intensity), paying particular 
attention to correctly reproducing the strength of some critical lines, 
like the [OIII]$\lambda$5007\AA\ and the H$\alpha$.
We find that two sources 
(2XMMpJ145442.2+182937 
and 
2XMMpJ115121.8$-$283605) are affected by a relatively high absorption (A$_{\rm V}>$ 2~mag 
and 
A$_{\rm V}\sim$ 2~mag), while the third (2XMMpJ151612.2+070341) is only moderately 
absorbed 
(A$_V\sim$1 mag). 
When translated into a value of N$_{\rm H}$ using the standard Galactic  relation
(e.g.~\cite{maiolino-a}), the absorption levels inferred from the optical 
spectra are 
N$_{\rm H} \, \llappr (2-4) \, \times$10$^{21}$~cm$^{-2}$, i.e.~lower
(significantly lower,
in the case of 2XMMpJ151612.2+070341) than the values obtained from the X-ray data 
(see \S\ref{par3}). 

The existence of a fraction of AGN where the level of absorption deduced by 
the X-ray spectra
is significantly higher than what can be inferred from the optical data is well known 
and discussed 
in the literature (e.g.~\cite{tommaso}; \cite{maiolino-b}; \cite{max}). To date, we 
are not in a
position 
of quantifying the frequency of this type of source since the sources 
discussed here do not
constitute a complete, well-defined sample of objects. It is very likely that 
the way we have 
chosen
these objects out of the 2XMMp catalogue,
in particular the requirement of a bright optical counterpart,
has favoured the selection of objects that are highly absorbed 
in the X-ray, but with a low/moderate optical absorption. 
A larger and unbiased sample of HR-selected 
sources would be instrumental in evaluating the fraction of sources strongly
absorbed in the
X-rays but only moderately absorbed in the optical band.

\section{Concluding remarks}
\label{par6}

Three sources of the 2XMMp catalogue with hard X-ray colour 
strongly suggesting the presence of absorption
were also observed at NTT@ESO.
All three sources show high absorption (N$_{\rm H}>10^{22}$~cm$^{-2}$)
in the X-ray band and moderate absorption (A$_{\rm V} \sim 1-2$ ~mag) in the optical. 
We can therefore definitely state that we are dealing with a
successful selection criterion able to ``isolate'' obscured sources in X-rays.
\cite{page} have been studying hard-spectrum XMM-{\em Newton} sources with 
2$-$4.5~keV X-ray flux $>$ $10^{-14}$ erg cm$^{-2}$ s$^{-1}$ and
get the same high efficiency in finding absorbed X-ray sources
(6.3 $\times 10^{21} <$ N$_{\rm H}<2.5 \times 10^{23}$~cm$^{-2}$).

We note that, of the large number (427) of high-latitude sources with HR3 $> 0$ 
and ML $> 20$
available in the 2XMM pre-release Serendipitous EPIC Source Catalogue, 
381 have a MOS2 X-ray flux in the 4.5$-$12~keV band larger than $10^{-13}$ erg cm$^{-2}$ s$^{-1}$.
This means that their optical spectroscopy should be achievable with ground-based telescopes,
as already happened for most of the sources of the HBS.
We can therefore foresee a valuable increment to the known sample of absorbed AGN, 
up to {\em a few hundred} newly absorbed AGN, and about 1/3
among them, as inferred from the HBS statistics, should be 
the rare type~2 quasars. 



\begin{acknowledgements}
We acknowledge financial contribution from the contract ASI-INAF I/023/05/0. 
Part of this work was supported by a Cofin grant for the year 2006.
\end{acknowledgements}

\end{document}